\documentclass[prl,showpacs,showkeys,twocolumn]{revtex4}%
\usepackage{amsfonts}
\usepackage{graphicx}
\usepackage{amsmath}
\usepackage{amssymb}%
\begin{document}
\preprint{UH, UCM}
\title[Quasiclassical frustration]{Quasiclassical frustration}
\author{H. Kohler}
\affiliation{Institut f\"{u}r theoretische Physik, Philosophenweg 19, Universit\"{a}t
Heidelberg, D-69120 Heidelberg, Germany}
\email{kohler@tphys.uni-heidelberg.de}
\author{F. Sols}
\affiliation{Departamento de F\'{\i}sica de Materiales, Facultad de Ciencias F\'{\i}sicas,
Universidad Complutense de Madrid, E-28040 Madrid, Spain}
\email{f.sols@fis.ucm.es}

\begin{abstract}
We study the dissipative properties of a harmonic oscillator subject to two
independent heat baths, one of which couples to its position and the other one
to its momentum. This model describes a large spin impurity in a ferromagnet.
We find that some effects of the two heat baths partially cancel each other.
Most notably, oscillations may remain underdamped for arbitrarily strong
coupling. This effect is a direct consequence of the mutually conjugate
character of position and momentum. For a single dissipative bath coupled
linearly to both position and momentum, no underdamped regime is possible for
strong coupling. The dynamics of purity loss for one and two wave packets is
also investigated.

\end{abstract}
\volumeyear{year}
\volumenumber{number}
\issuenumber{number}
\eid{identifier}
\startpage{1}
\endpage{ }
\maketitle

The harmonic oscillator provides a scenario where the physics of a quantum
particle coupled to a heat bath can be investigated analytically
\cite{ulle66,weis99,cald84}. Thus its study may shed light on physical effects
that have been found in less tractable models. Recently, Castro Neto
\textit{et al.} \cite{cast03} have investigated, by means of perturbative
renomalization group methods, the equilibrium dynamics of a quantum magnetic
impurity in a ferromagnet and have found that the symmetric coupling of the
magnon bath to the $x$ and $y$ spin components ($z$ being the magnetization
direction) effectively decreases the strength of the coupling. They have
coined this term \textit{quantum frustration}, since it may be interpreted as
the inability of the spin bath to simultaneously measure two non-commuting
observables such as the $x$ and $y$ components of the impurity spin. An
important consequence is that spin coherence may be longer lived than it would
if only one of the spin components were coupled to the dissipative bath. This
effect could be relevant for quantum information, as it would provide a
mechanism for quantum spins to remain coherent for long times.

Here we investigate the effect of quantum frustration in a different but
exactly tractable physical system, namely, a harmonically oscillator coupled
separately, through its position and momentum, to two independent oscillator
heat baths \cite{cald83}. In the symmetric limit, this model describes the
behavior of a large spin magnetic impurity in a ferromagnet. Another instance
of a physical system which interacts, through position and momentum, with two
different baths is a Josephson junction, where the relative particle number
(proportional to the electric dipole) couples to the radiation field while the
relative phase interacts with the quasiparticle field \cite{kohl04}. Those two
environments have different spectral properties. To investigate a more
symmetric coupling, we consider a harmonic oscillator whose position and
momentum interact linearly with two independent Ohmic baths \cite{legg84}. We
find that the two baths do cancel in some but no all respects. A degree of
cancellation is revealed by the persistence of underdamped oscillations for
arbitrarily strong dissipation provided that the two baths couple with
comparable strength. Quantum purity is weakened by the presence of a symmetric
second bath although its decay is slowed down. Since a symmetrically damped
harmonic oscillator behaves like a large spin in a ferromagnet, we refer to
such a mixed situation as \textit{quasiclassical frustration}.

We investigate the following model Hamiltonian \cite{kohl04}:
\begin{equation}
H=V_{q}(q+\delta q)+V_{p}(p+\delta p)+\sum_{k}\omega_{qk}a_{qk}^{\dagger
}a_{qk}+\sum_{k}\omega_{pk}a_{pk}^{\dagger}a_{pk}\ , \label{mod1}%
\end{equation}
\begin{align}
\delta q  &  =ig_{p}(p+\delta p)\sum_{k}C_{pk}\left(  a_{pk}^{\dagger}%
-a_{pk}\right)  \ \nonumber\\
\delta p  &  =ig_{q}(q+\delta q)\sum_{k}C_{qk}\left(  a_{qk}^{\dagger}%
-a_{qk}\right)  ~, \label{mod2}%
\end{align}
where $[q,p]=i$ and\ $g_{q},g_{p}$ are sufficiently well behaved functions.
For general $g_{q},g_{p}$ the two equations in Eq.~(\ref{mod2}) cannot be
decoupled without generating interactions between the two baths. However, if
one of the coupling functions, say $g_{p}$, equals $1$, then $[\delta p,\delta
q]=0$ and it becomes possible to remove both fluctuating contributions from
the potential terms through the unitary transformations $U_{p}=\exp(ip\delta
q)$ and $U_{q}=\exp(i\int^{q+\delta q}\delta p(q^{\prime})dq^{\prime})$. One
arrives at the Hamiltonian
\begin{align}
H  &  =V_{q}(q)+\sum_{k}\omega_{qk}\left\vert a_{qk}+\frac{C_{qk}}{\omega
_{qk}}\int^{q}dq^{\prime}g_{q}(q^{\prime})\right\vert ^{2}\nonumber\\
&  +V_{p}(p)+\sum_{k}\omega_{pk}\left\vert a_{pk}+\frac{C_{pk}}{\omega_{qk}%
}p\right\vert ^{2}\ , \label{compact}%
\end{align}
where the short-hand notation $|a|^{2}\equiv a^{\dagger}a$ has been used.

At this point it is important to specify what we mean by coupling to position
or momentum. Those are the particle variables to which the environment couples
as a set of otherwise independent harmonic oscillators. This is the case e.g.
in (\ref{compact}), where the bath oscillators would remain independent if $q$
and $p$ were c-numbers. On the contrary, this is not the case in (\ref{mod1}),
where, due to the nonlinear character of $V_{q}$ and $V_{p}$, the bath
oscillators do interact with each other \cite{comm5}. A popular example is
that of a charged particle interacting with the photon field \cite{kohl04}.
When the velocity-coupling model is adopted, it must be accompanied by a
diamagnetic term that contains an interaction between photons. By contrast,
through a canonical transformation, the photon field interacts with the
particle position without an explicit interphoton interaction. Thus, within
the precise convention we propose here, the electromagnetic field couples to
the position of a charged particle.

We focus on the case of a harmonic oscillator [$V_{q}(q)=\omega_{q}q^{2}/2$,
$V_{p}(p)=\omega_{p}p^{2}/2$] with $g_{q}=g_{p}=1$:%
\begin{align}
H &  =\frac{\omega_{q}}{2}q^{2}+\sum_{k}\omega_{qk}\left\vert a_{qk}%
+\frac{C_{qk}}{\omega_{qk}}q\right\vert ^{2}\nonumber\\
&  +\frac{\omega_{p}}{2}p^{2}+\sum_{k}\omega_{pk}\left\vert a_{pk}%
+\frac{C_{pk}}{\omega_{pk}}p\right\vert ^{2}~.\label{gugu}%
\end{align}
This is the model of dissipative coupling that displays the highest degree of
symmetry between $q$ and $p$. Hence it is interesting to study frustration in
a physical system other than a quantum spin in a ferromagnet. The two baths of
independent harmonic oscillators are described by the spectral densities
\begin{equation}
J_{n}(\omega)\ =\ 2\sum_{k}|C_{nk}|^{2}\delta(\omega-\omega_{nk}%
)\ ,~~~~~~~n=p,q\ .\qquad\label{spectra}%
\end{equation}
We as\-sume a power law be\-havior at $\omega=0$ and write $J_{n}(\omega
)$$=2\gamma_{n}\omega^{\alpha_{n}}$$/(\omega_{\mathrm{ph}}^{\alpha_{n}-1}\pi
)$, where the introduction of the frequency $\omega_{\mathrm{ph}}$ renders the
coupling constants $\gamma_{n}$ dimensionless. Moreover, large cutoff
frequencies $\Omega_{n}$ are assumed to exist for both environments.

Eliminating the bath variables, the Heisenberg equations of motion for $q$ and
$p$ are obtained,
\begin{align}
\dot{q}(t)  &  =\omega_{p}p(t)+\int^{t}ds\,K_{p}(t-s)\dot{p}(s)+F_{p}%
(t)\nonumber\\
-\dot{p}(t)  &  =\omega_{q}q(t)+\int^{t}ds\,K_{q}(t-s)\dot{q}(s)+F_{q}(t),
\label{equations of motion}%
\end{align}
where $K_{n}(t)\equiv\int_{0}^{\infty}J_{n}(\omega)\cos(\omega t)d\ln\omega$
and $F_{n}(t)=\sum_{k}C_{nk}a_{nk}\exp(-i\omega_{nk}t)+\mathrm{H.c}$. In
Fourier space, Eq. (\ref{equations of motion}) reads%
\begin{align}
\left[  \widetilde{J}_{q}(\omega)-\omega_{q}\right]  q+i\omega p  &
=F_{q}\label{sist-Fourier}\\
-i\omega q+\left[  \widetilde{J}_{p}(\omega)-\omega_{p}\right]  p  &  =F_{p}~,
\end{align}
where $\widetilde{J}_{n}(\omega)$ is the symmetrized Riemann transform
\cite{abr72}:
\begin{equation}
\widetilde{f}(\omega)\ \ =\ \omega^{2}\mathcal{P}\int_{0}^{\infty}%
\frac{f(\omega^{\prime})}{\omega^{\prime}\left(  {\omega^{\prime}}^{2}%
-\omega^{2}\right)  }d\omega^{\prime}-i\mathrm{sgn\,}(\omega)\frac{\pi}%
{2}f(|\omega|)\quad. \label{tildetransformed}%
\end{equation}

The oscillation modes are given by the zeros of the polynomial%
\begin{equation}
\chi^{-1}(\omega)=\omega_{0}^{2}-\omega^{2}-\omega_{q}\widetilde{J}_{p}%
(\omega)-\omega_{p}\widetilde{J}_{q}(\omega)+\widetilde{J}_{q}(\omega
)\widetilde{J}_{p}(\omega) \label{ee13}%
\end{equation}
where $\chi(\omega)$ is the generalized susceptibility.

\begin{figure}
[ptb]
\begin{center}
\includegraphics[
trim=0.000000in 0.000000in 0.259785in 0.000000in,
height=2.3603in,
width=2.3611in
]%
{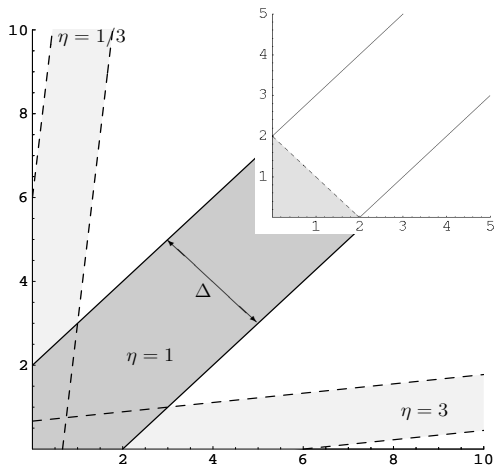}%
\caption{Stripes of underdamped oscillations in the $(\gamma_{q},\gamma_{p})$
plane for three different values of the parameter $\eta=\left(  \omega
_{q}/\omega_{p}\right)  ^{1/2}$, for $q$ and $p$ coupled to different baths.
Inset: Shaded region denotes underdamping for a single bath coupled to $q$ and
$p$.}%
\label{stripes}%
\end{center}
\end{figure}

We further assume that the two heat baths are Ohmic: $J_{n}(\omega)$
$=2\gamma_{n}\omega$ $/\pi$. For $\Omega_{n}\rightarrow\infty$, this implies
$\widetilde{J}_{n}(\omega)\rightarrow i\gamma_{n}\omega$. Then the
eigenfrequencies are given by%
\begin{equation}
\omega_{0}^{2}-i(\omega_{q}\gamma_{p}+\omega_{p}\gamma_{q})\omega
-(1+\gamma_{q}\gamma_{p})\omega^{2}=0~, \label{second-degree}%
\end{equation}
the solutions being
\begin{equation}
\omega_{\pm}=\frac{\omega_{0}}{\left(  1+\gamma_{q}\gamma_{p}\right)  ^{1/2}%
}\left(  -i\kappa\pm\sqrt{1-\kappa^{2}}\right)  , \label{solu}%
\end{equation}%
\begin{equation}
\kappa\equiv\frac{\gamma_{q}\omega_{p}+\gamma_{p}\omega_{q}}{2\omega
_{0}\left(  1+\gamma_{q}\gamma_{p}\right)  ^{1/2}}\ . \label{kappa}%
\end{equation}
The transition from $\kappa<1$ to $\kappa>1$ marks the crossover from
underdamped to overdamped oscillations. The condition $\kappa<1$ requires
(criterion A)
\begin{equation}
|\gamma_{q}\omega_{p}-\gamma_{q}\omega_{p}|<2\omega_{0}
\label{general-condition-A}%
\end{equation}
The underdamped region satisfying (\ref{general-condition-A}) lies in a stripe
of width $\Delta=4\eta\left(  1+\eta^{4}\right)  ^{-1/2}$ with $\eta=\left(
\omega_{q}/\omega_{p}\right)  ^{1/2}$, limited by the graphs of the functions
$f(\gamma_{q})=\ (\gamma_{q}\pm2\eta)\eta^{-2}$. The stripes of underdamped
oscillations in the $(\gamma_{q},\gamma_{p})$ plane are plotted in
Fig.~\ref{stripes} for $\eta=1/3,1,3$. A remarkable consequence is that, given
a value of e.g. $\gamma_{q}$, one may drive the system from the overdamped to
the underdamped regime by \textit{increasing} $\gamma_{p}$.%
For $\eta=1$, the oscillator is underdamped if $\gamma_{q}\simeq\gamma_{p}$,
i.e., if the couplings to the two baths are of comparable strength. When
$\gamma_{q}=\gamma_{p}\equiv\gamma$ the dimensionless parameter $\kappa$
becomes,%
\begin{equation}
\kappa=\gamma\left(  1+\gamma^{2}\right)  ^{-1/2}<1~\ .
\label{kappa-symmetric}%
\end{equation}
Thus, in the fully symmetric case, the oscillator remains underdamped
for\textit{ all }values of the coupling strength\textit{. }This is in contrast
with the case of one noise ($\gamma_{p}=0$), which requires $\gamma_{q}<2\eta$
to be underdamped. Conversely, if $\gamma_{q}=0$, the condition is $\gamma
_{p}<2\eta^{-1}$. The inset of Fig. \ref{stripes} \ shows the underdamped
region for an oscillator coupling through $q$ and $p$ to a \emph{single} heat
bath \cite{kohl05}. Surprisingly, the behavior is qualitatively different in
that underdamping is always lost for sufficiently large coupling.

Another interesting quantity is $D_{q}(\omega)\equiv\operatorname{Im}\chi
_{qq}(\omega)/\omega$, where $\chi_{qq}(\omega)$ is the Fourier transform of
$\langle\lbrack q(t),q(0)]\rangle$. For Ohmic environments,
\begin{equation}
D_{q}(\omega)=\frac{\gamma_{q}\omega_{p}^{2}+\gamma_{p}(1+\gamma_{q}\gamma
_{p})\omega^{2}}{[(1+\gamma_{q}\gamma_{p})\omega^{2}-\omega_{0}^{2}%
]^{2}+(\gamma_{q}\omega_{p}+\gamma_{p}\omega_{q})^{2}\omega^{2}}~.
\label{corr-function}%
\end{equation}
Following Ref. \cite{cast03}, we may view the presence of a peak in
$D_{q}(\omega)$ as a signature for the existence of coherent dynamics
(criterion B). This occurs for%
\begin{equation}
\gamma_{q}^{3}\omega_{p}^{4}<\left(  2\gamma_{q}\omega_{p}^{2}+\gamma
_{p}\omega_{0}^{2}\right)  \omega_{0}^{2}.
\end{equation}
In the symmetric case, this translates into $\gamma<\sqrt{3}$. For an
oscillator coupled to a single bath through its position ($\gamma_{p}=0$), the
requirement is $\gamma_{q}<\sqrt{2}\eta$. \ Finally, if $\gamma_{q}=0$,
$D_{q}(\omega)$ always displays a peak \cite{comm2}.

A third possible condition for the existence of coherent dynamics is that, in
Eq. (\ref{solu}),%
\begin{equation}
|\operatorname{Im}\omega_{\pm}|<|\operatorname{Re}\omega_{\pm}|~.
\label{condition C}%
\end{equation}
(criterion C). This yields%
\begin{equation}
\gamma_{q}^{2}\omega_{p}^{2}+\gamma_{p}^{2}\omega_{q}^{2}<2\omega_{0}^{2}~.
\label{general-condition-C}%
\end{equation}
In the symmetric case, the condition (\ref{general-condition-C}) becomes
$\gamma<1$. For $\gamma_{p}=0$, it becomes $\gamma_{q}<\sqrt{2}\eta$, while
for $\gamma_{q}=0$, it reads $\gamma_{p}<\sqrt{2}\eta^{-1}$.

In table \ref{TableKey-5}, the coherence signatures for the main three
particular cases are summarized. Criteria A and C are symmetric in $q$ and
$p$, but not criterion B. The general trend (particularly clear if one
considers A and B) is that, starting from a single dissipative bath coupled to
e.g. $q$, the introduction of a second bath that couples to $p$ with the same
spectrum and comparable strength favors coherent and underdamped dynamics. For
example, an oscillator with $\eta=1$ that is driven from $\gamma_{p}=0$ to
$\gamma_{p}=\gamma_{q}=\gamma$, with $\gamma_{q}$ fixed at a value $\sqrt
{2}<\gamma_{q}<\sqrt{3}$, will cross over from incoherent to coherent behavior
under both criteria A and B.

We have noted the striking result that, in the symmetric case, the oscillator
is underdamped for all values of $\gamma$. However, criteria B and C indicate
that the oscillator lies deep in the underdamped region only if $\gamma$ is
small. Such a limitation is also patent in the (possible) maximum of
$D_{q}(\omega)/D_{q}(0)$ as well as in the ratio $|\operatorname{Re}%
\omega_{\pm}|/|\operatorname{Im}\omega_{\pm}|=\gamma^{-1}$. Both quantities
stay well above unity only if $\gamma$ is small. A related point is that, as
$\gamma\rightarrow\infty$, the ratio $D_{q}(\omega)/D_{q}(0)$ does not
saturate but rather decays as $\omega_{0}^{2}/\gamma^{2}\omega^{2}$ for
nonzero $\omega$.%

\begin{table}[tbp] \centering
\begin{tabular}
[c]{cc|c|c|c}
&  & symmetric & only $\gamma_{q}$ & only $\gamma_{p}$\\\hline
$\kappa<1$ & (A) & \textit{always} & $\gamma_{q}<2\eta$ & $\gamma_{p}%
<2\eta^{-1}$\\\hline
$D_{q}(\omega)$ has peak & (B) & $\gamma<\sqrt{3}$ & $\gamma_{q}<\sqrt{2}\eta$
& \textit{always}\\\hline
$|\operatorname{Im}\omega_{\pm}|<|\operatorname{Re}\omega_{\pm}|$ & (C) &
$\gamma<1$ & $\gamma_{q}<\sqrt{2}\eta$ & $\gamma_{p}<\sqrt{2}\eta^{-1}$%
\end{tabular}
\caption{Condition for coherent dynamics according to three different criteria (left column, labeled A, B, C), for three
particular cases (upper row). The symmetric limit includes the assumption $\eta=1$. General case is given in main text. }\label{TableKey-5}%
\end{table}%

An oscillator initially prepared in a pure coherent state that at $t=0$ begins
to interact linearly with an oscillator bath is described by a reduced density
$\rho$ which remains Gaussian at all times. Then the purity $\mathcal{P}%
(t)\equiv\operatorname{Tr}(\rho^{2})$ is given by%
\begin{equation}
\mathcal{\mathcal{P}}^{-2}(t)=4\langle q^{2}\rangle\langle p^{2}\rangle~.
\label{general-purity}%
\end{equation}

At long times, $\langle q^{2}\rangle$ and $\langle p^{2}\rangle$ reach their
equilibrium values, which contain contributions from both baths as well as
hybrid terms which vanish if any of the two baths disappears \cite{kohl04}.
For simplicity, we focus on the zero temperature, weak coupling case. Then,%
\begin{equation}
\langle q^{2}\rangle=\frac{1}{2\eta}-\frac{\gamma_{q}}{2\eta^{2}}+\gamma
_{p}\left(  \ln\frac{\Omega_{p}}{\omega_{0}}-\frac{1}{2}\right)
+\mathcal{O}(\gamma_{q},\gamma_{p})\ ,\label{meanqpaper}%
\end{equation}
and similarly for $\langle p^{2}\rangle$. The two baths have opposite effects
on $\langle q^{2}\rangle$, but for $\Omega_{n}\gg\omega_{0}$ the logarithmic
divergence from the $p$-coupled bath overwhelms the squeezing of $q$ favored
by the $q$-coupled environment. An impure mixture results:%
\begin{align}
\mathcal{\mathcal{P}}_\infty^{-2} &  \simeq1+\frac{2\gamma_{q}}{\eta}\left(
\ln\frac{\Omega_{p}}{\omega_{0}}-1\right)  \nonumber\\
&  +2\gamma_{p}\eta\left(  \ln\frac{\Omega_{q}}{\omega_{0}}-1\right)
+\mathcal{O}(\gamma_{q},\gamma_{p})\gg1.\label{product}%
\end{align}

The dynamical evolution of the purity $\mathcal{P}(t)$ is cumbersome in the
general case, but it becomes tractable in the symmetric problem. We find
($\Omega_{q}=\Omega_{p}=\Omega$)
\begin{equation}
\mathcal{P}(t)\simeq\left\{
\begin{tabular}
[c]{cc}%
$e^{-\Omega t}$~, & $0\leq t\lesssim\Omega^{-1}$\\
$\mathcal{\mathcal{P}}_\infty\left[  1+\frac{2\gamma}{\left(  1+\gamma
^{2}\right)  ^{3/2}}\frac{e^{-t/\tau}}{t^{2}}\right]  ,$ & $\Omega^{-1}\ll
t\rightarrow\infty~,$%
\end{tabular}
\ \ \ \ \right.  \label{purityresult}%
\end{equation}
with $\tau^{-1}=\gamma\omega_{0}/(1+\gamma^{2})=\left\vert \operatorname{Im}%
\omega_{\pm}\right\vert $. The fast decay on the scale of $\Omega^{-1}$ comes
from the choice of decoupled initial conditions \cite{sanc94} and from the
concurrence of two baths \cite{comm9}. For $t\gg\Omega^{-1}$, the system
evolves slowly towards equilibrium. The divergence of $\tau$ for
$\gamma\rightarrow\infty$ might be interpreted as robustness against purity
loss. However, it should be noted that such a slowing down merely reflects a
dilation of all time scales with increased friction. For instance, $\left\vert
\operatorname{Re}\omega_{\pm}\right\vert =\omega_{0}/(1+\gamma^{2})$ vanishes
even faster.

We have also investigated the purity decay when at $t=0$ the system is
prepared in a linear superposition of two coherent states centered at $q=\pm
a/2$ with zero average momentum \cite{comm7}. For the symmetric case, we find
\cite{kohl05}%
\begin{equation}
P(t)=\frac{\mathcal{P}(t)}{2}\left\{  1+\frac{\cosh^{2}\left[  \frac{a^{2}}%
{4}(\phi(t)\mathcal{P}(t)-\frac{1}{2})\right]  }{\cosh^{2}\left(
a^{2}/8\right)  }\right\}  , \label{two-purity}%
\end{equation}
where $\phi(t)$ is a complicated function that evolves from $\phi(0)=1$ to
$\lim_{t\rightarrow\infty}\phi(t)=0$, and the single wave packet purity
$\mathcal{P}(t)$ is given in (\ref{purityresult}). As expected,
$P(t)/\mathcal{P}(t)\rightarrow1$ as $a\rightarrow0$, and $P(t)/\mathcal{P}%
(t)\rightarrow1/2$ as $a\rightarrow\infty$.

Interestingly, the structure of (\ref{two-purity}) is such that, as time
passes and $\phi(t)\mathcal{P}(t)$ evolves from $1$ to $\ 0$, the ratio
$P(t)/\mathcal{P}(t)$ starts at unity, as corresponds to a pure state, then
decreases and finally, at long times, goes back to unity. 
When $a$ is large $P(t)/\mathcal{P}(t)$ decays rapidly on a timescale $\sim 1/4a^2\gamma$ 
to $1/2$. There it stays for a time which increases with distance as $\sim \gamma^{-1}\ln a$.
Afterwards it returns to one.    
The ratio 1/2 can be rightly interpreted as resulting
from the incoherent mixture of the two wave packets. Thus it comes as a
relative surprise that $P(t)/\mathcal{P}(t)$ becomes unity again at long
times, as if coherence among the two wave packets were eventually recovered.
The physical explanation lies in the ergodic character of the long time
evolution, with both wave packets evolving towards the equilibrium
configuration described in Eqs. \ref{general-purity}-\ref{product}. Once the
two initially separate wave packets begin to overlap, they regain mutual
coherence. Due to the symmetry of \ the problem, a similar result would have
been obtained if the oscillator had started from a superposition of two
coherent states located in the same region of real space but with different
average values of the momentum.

In summary, we have found features that are reminiscent of an effective
particle-bath decoupling, such as the persistence of underdamped oscillations
for arbitrarily large values of $\gamma$ in the case of a symmetric oscillator
and the slowing down of purity decay for a Gaussian wavepacket. Another
feature is that two initially separate wave packets regain relative coherence
at long times because they recombine.
The situation is reminiscent of the quantum frustration exhibited by a
magnetic quantum impurity albeit in a more limited form \cite{cast03}. The
main difference between the two problems is the dimensionality of the particle
Hilbert space. Evolving in the continuum, the quantum oscillator can be
considerably degraded by the effect of the environment, as shown in
(\ref{purityresult}). It is only when the two wave packets recombine because
of ergodicity that relative purity is recovered. By contrast, the spin-1/2
magnetic impurity lives in a two-dimensional space. The only possible effect
of the environment is to flip the spin. Thus, in any representation, two
initially orthogonal states quickly overlap and tend to preserve mutual
coherence. The net result is an increased decoupling from a symmetrically
dissipative environment. The requirement of low dimensionality suggests that
strong frustration is a genuinely quantum effect.

We would like to thank F. Guinea and A. J. Leggett for valuable discussions.
This research has been supported by MEC (Spain) under Grants No. BFM2001-0172
and FIS2004-05120, the EU RTN Programme under Contract No. HPRN-CT-2000-00144,
and the Ram\'{o}n Areces Foundation.


\begin{thebibliography}{99}
\bibitem {comm5}An example is the electromagnetic field coupled to a charge
position \cite{kohl04}. The alternative velocity-coupling model requires a
diamagnetic term which is quadratic in the photon field.
\bibitem {comm2}This case resembles (but is not identical) to that of a
particle coupled through its position to an environment with $J_{q}%
(\omega)\propto\omega^{3}$ \cite{kohl04}.
\bibitem {comm9}For $\gamma_{p}=0$, the time scale of relaxation is
$\gamma_{q}^{-1}$  and $\mathcal{P}(t)$ has no initial slip.
\bibitem {comm7}The decay of coherence of such a system in the presence of one
noise has been studied in Ref. \cite{cald84} but without explicit attention to
the purity evolution.
\end{thebibliography}
\end{document}